\documentclass[10pt]{article}
\usepackage[final]{graphics}
\usepackage{amsmath}
\usepackage{amssymb}
\usepackage{amsfonts,amsbsy}

\def\empile#1\over#2{\mathrel{\mathop{\kern 0pt#1}\limits_{#2}}}

\newcommand{\sll}{\raise.15ex\hbox{$/$}\kern-.43em\hbox{$l$}}
\newcommand{\slepsilon}{\raise.15ex\hbox{$/$}\kern-.53em\hbox{$\epsilon$}}
\newcommand{\slvarepsilon}{\raise.15ex\hbox{$/$}\kern-.53em\hbox{$\varepsilon$}}
\newcommand{\slL}{\raise.15ex\hbox{$/$}\kern-.53em\hbox{$L$}}
\newcommand{\slP}{\raise.15ex\hbox{$/$}\kern-.53em\hbox{$P$}}
\newcommand{\slp}{\raise.1ex\hbox{$/$}\kern-.63em\hbox{$p$}}
\newcommand{\slq}{\raise.1ex\hbox{$/$}\kern-.53em\hbox{$q$}}
\newcommand{\slv}{\raise.1ex\hbox{$/$}\kern-.63em\hbox{$v$}}
\newcommand{\slR}{\raise.15ex\hbox{$/$}\kern-.53em\hbox{$R$}}
\newcommand{\slQ}{\raise.15ex\hbox{$/$}\kern-.53em\hbox{$Q$}}
\newcommand{\slK}{\raise.15ex\hbox{$/$}\kern-.53em\hbox{$K$}}
\newcommand{\slk}{\raise.15ex\hbox{$/$}\kern-.53em\hbox{$k$}}
\newcommand{\slSigma}{\raise.15ex\hbox{$/$}\kern-.53em\hbox{$\Sigma$}}
\newcommand{\slcalP}{\raise.15ex\hbox{$/$}\kern-.63em\hbox{$\cal P$}}
\newcommand{\slA}{\raise.15ex\hbox{$/$}\kern-.73em\hbox{$A$}}
\newcommand{\slbfA}{\raise.15ex\hbox{$/$}\kern-.73em\hbox{${\imb A}$}}
\newcommand{\slpartial}{\raise.15ex\hbox{$/$}\kern-.53em\hbox{$\partial$}}
\newcommand{\sla}{\raise.15ex\hbox{$/$}\kern-.53em\hbox{$a$}}
\newcommand{\slb}{\raise.15ex\hbox{$/$}\kern-.53em\hbox{$b$}}
\newcommand{\slc}{\raise.15ex\hbox{$/$}\kern-.53em\hbox{$c$}}
\newcommand{\slD}{\raise.15ex\hbox{$/$}\kern-.53em\hbox{$D$}}
\newcommand{\slC}{\raise.15ex\hbox{$/$}\kern-.53em\hbox{$C$}}

\def\p{{\boldsymbol p}}

\def\x{{\boldsymbol x}}
\def\y{{\boldsymbol y}}
\def\X{{\boldsymbol X}}

\def\r{{\boldsymbol r}}
\def\z{{\boldsymbol z}}

\catcode`\@=11

\newcount\@tempcntc
\def\@citex[#1]#2{\if@filesw\immediate\write\@auxout{\string\citation{#2}}\fi
    \@tempcnta\z@\@tempcntb\m@ne\def\@citea{}\@cite{%
          \@for\@citeb:=#2\do%
      {\@ifundefined{b@\@citeb}%
          {\@citeo\@tempcntb\m@ne\@citea%
                  \def\@citea{,\penalty\@m\ }{\bf ?}\@warning%
                  {Citation `\@citeb' on page \thepage \space undefined}}%
          {\setbox\z@\hbox{\global\@tempcntc0\csname 
b@\@citeb\endcsname\relax}
       \ifnum\@tempcntc=\z@ \@citeo\@tempcntb\m@ne%
         \@citea\def\@citea{,\penalty\@m}%
         \hbox{\csname b@\@citeb\endcsname}%
       \else%
        \advance\@tempcntb\@ne%
        \ifnum\@tempcntb=\@tempcntc%
        \else\advance\@tempcntb\m@ne\@citeo%
        \@tempcnta\@tempcntc\@tempcntb\@tempcntc\fi\fi}}\@citeo}{#1}}%

\def\@citeo{\ifnum\@tempcnta>\@tempcntb\else\@citea
    \def\@citea{,\penalty\@m}%
    \ifnum\@tempcnta=\@tempcntb\the\@tempcnta\else
     {\advance\@tempcnta\@ne\ifnum\@tempcnta=\@tempcntb \else
\def\@citea{--}\fi
      \advance\@tempcnta\m@ne\the\@tempcnta\@citea\the\@tempcntb}\fi\fi}

\catcode`\@=12

\begin{document}

\thispagestyle{empty}
\title {\bf Searching  Evidence for the Color Glass Condensate at RHIC}

\author{Jean-Paul Blaizot, Fran\c cois Gelis}
\maketitle
\begin{center}
Service de Physique Th\'eorique\footnote{URA 2306 du CNRS.}\\
    B\^at. 774, CEA/DSM/Saclay\\
    91191, Gif-sur-Yvette Cedex, France
\end{center}

\begin{abstract}
   This contribution discusses the phenomenon of parton saturation, the
   color glass picture of hadronic wavefuntions, and their relevance in
   the early stages of nucleus-nucleus collisions. Evidence for the
   color glass condensate in the presently available RHIC data is
   critically reviewed.
\end{abstract}
\vskip 5mm
\begin{flushright}
Preprint SPhT-T04/062
\end{flushright}

\section{Introduction}

The degrees of freedom involved in the early stages of a
nucleus-nucleus collision at sufficiently high energy are partons,
mostly gluons, whose density grows as the energy increases (i.e., when
$x$, their momentum fraction, decreases). This growth of the number of
gluons in the hadronic wave functions is a phenomenon which has been
well established at HERA. One expects however that it should
eventually ``saturate'' when non linear QCD effects start to play a
role.

The existence of such a saturation regime has been predicted long ago,
together with estimates for the typical transverse momenta where it
sets in. But it is only during the last decade that equations
providing a complete dynamical description of the saturated regime
have been obtained. A remarkable feature which emerges from the
solution of these equations is that the dense, saturated system of
partons to be found in hadronic wave functions at high energy has
universal properties, the same for all hadrons or nuclei. It follows
that the early stages of hadronic collisions at sufficiently high
energies are governed by universal wave functions (``color glass
condensate''), whose properties could in principle be calculated from
QCD. This is a very exciting perspective which fully justifies the
active search for evidence of this novel regime of QCD both at HERA
and at RHIC.

It is expected that the saturation regime sets in earlier (i.e., at
lower energy) in collisions involving large nuclei than in those
involving protons. In fact, the parton densities in the central
rapidity region of a Au-Au collision at RHIC are not too different
from those measured in deep inelastic scattering at HERA. There is
however one important difference: while at HERA these densities result
from gluon evolution, at RHIC there is little evolution, at least in
the central rapidity region, and the large densities result mostly
from the additive contributions of the participant nucleons. Of
course, one has the possibility at RHIC to explore various situations.
In particular the study of $dA$ collisions in the fragmentation region
of the deuteron gives access to a regime where final state
interactions should play a minor role and where quantum evolution
could be significant. Indeed, very exciting results have been obtained
in this regime.

One may worry that at present energies, the conditions for saturation
are at best marginally satisfied.  Nevertheless a successful
phenomenology based on the saturation picture has been developed over
the last few years, both at RHIC and at HERA. We shall review in this
report some of the experimental findings providing evidence for the
color glass picture.  It should be emphasized that the field is still
in a rapidly evolving phase, most theoretical analysis are incomplete,
so that conclusions reached today can at best be considered as
tentative. Before we go into this phenomenological discussion, we
shall start by a brief (and incomplete) historical perspective, to
emphasize the remarkable theoretical developments of the last decade.

\section{The infancy of the idea of saturation}

An important feature of partonic interactions is that they involve
mostly partons with comparable rapidities. Let us then consider a
nucleus-nucleus collision in the target rest frame and consider what
happens when one boosts the projectile, increasing its rapidity in
successive steps.  In the first step, the valence constituents become
Lorentz contracted in the longitudinal direction while the time scale
of their internal motions is Lorentz dilated.  In addition, the boost
reveals new vacuum fluctuations coupled to the boosted valence
partons.  Such fluctuations are not Lorentz contracted in the
longitudinal direction, and represents the dynamical degrees of
freedom.
Making an additional step in rapidity would freeze these
fluctuations, while making them Lorentz contracted as well. But the
additional boost also produces new quantum fluctuations, which become
the new dynamical variables. This argument can be repeated, and one
arrives at the picture of a high-energy projectile containing a large
number of frozen, Lorentz contracted partons (the valence partons,
plus all the quantum fluctuations produced in the previous boosts),
and partons which have a small rapidity and are not Lorentz contracted.
This space-time description was
developed before the advent of QCD (see for instance \cite{Feynm4}; in
Bjorken's lectures \cite{Bjork2}, one can actually foresee the modern
interpretation of parton evolution as a renormalization group
evolution {\it a la} Wilson on which the color glass formalism is
based).

Of course, such a space-time picture, which was deduced from rather
general field theoretical considerations, can now be understood in
terms of QCD. In fact, shortly after QCD was established as the theory
of strong interaction, quantitative equations were established,
describing the phenomenon outlined above
\cite{KuraeLF1,BalitL1,GriboL1,GriboL2,AltarP1,Doksh1}. In particular,
the equation derived by Balitsky, Fadin, Kuraev and Lipatov
\cite{KuraeLF1,BalitL1} describes the growth of the non-integrated
gluon distribution in a hadron as it is boosted towards higher
rapidities.
\begin{figure}[htbp]
\begin{center}
\resizebox*{!}{6cm}{\includegraphics{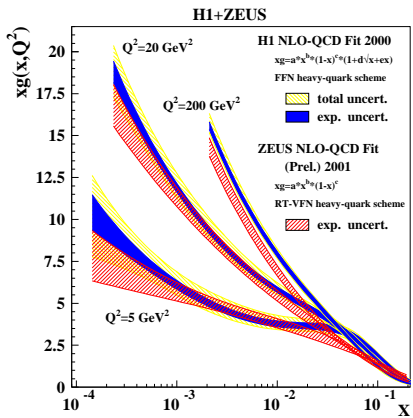}}
\end{center}
\caption{\label{fig:HERA-small-x} The gluon structure function in a
proton measured at HERA.}
\end{figure}
    Experimentally, an important increase of the number of gluons at
small $x$ has indeed been observed in the Deep Inelastic Scattering
experiments performed at HERA (see fig.~\ref{fig:HERA-small-x}), down
to $x\sim 10^{-4}$. Such a growth raises a problem: if it were to
continue to arbitrarily small $x$, it would induce an increase of
hadronic cross-sections as a power of the center of mass energy, in
violation of unitarity bounds.

However, as noticed by Gribov, Levin and Ryskin in \cite{GriboLR1},
the BFKL equation includes only branching processes that increase the
number of gluons ($g\to gg$ for instance), but not the recombination
processes that could reduce the number of gluons (like $gg\to g$).
While it may be legitimate to neglect the recombination process when
the gluon density is small, this cannot remain so at arbitrarily high
density: a saturation mechanism of some kind must set in.  Treating
the partons as ordinary particles, one can get a crude estimate of the
onset of saturation from a simple mean free path argument. The
recombination cross-section for gluon with transverse momentum $Q $ is
roughly given by
\begin{equation}
\sigma \sim \frac{\alpha_s(Q^2)}{Q^2}\; ,
\end{equation}
while the number of such gluons per unit of transverse
area is given by
\begin{equation}\label{rho}
\rho \sim \frac{xG(x,Q^2)}{\pi R^2}\; ,
\end{equation}
where $R$ is the radius of the hadron and $x$ the momentum fraction of
the considered gluons. Saturation sets in when $\rho\sigma\sim 1$, or
equivalently for:
\begin{equation}\label{Qsaturation}
Q^2= Q_s^2\; ,\quad{\rm with\ } Q_s^2
\sim \alpha_s(Q_s^2)\frac{xG(x,Q_s^2)}{\pi R^2}\; .
\end{equation}
The momentum scale that characterizes this new regime, $Q_s$, is
called the saturation momentum \cite{Muell3}. Partons with transverse
momentum $Q> Q_s$ are in a dilute regime; those with $Q<Q_s$ are in
the saturated regime.

A more refined argument for the onset of saturation was given in
\cite{MuellQ1}, where recombination is associated with a higher twist
correction to the DGLAP equation.  More generally, one may view the
saturation momentum $Q_s$ as the scale at which QCD non linear effects
become important.  This occurs when field fluctuations are such that
gluon kinetic energies become comparable to their interaction
energies, that is when $\langle(\partial A)^2\rangle\sim \alpha_s
\langle (A^2)^2\rangle$, where $\langle A^2 \rangle$ denotes the
fluctuations of the gauge fields. Since the relevant dynamics is in
the transverse plane, the magnitude of the gradient is fixed by the
transverse momentum $Q$. As for the magnitude of the gauge field
fluctuations, $\langle A^2 \rangle$, they can be estimated from the
particle number density in the transverse plane, i.e. $\langle
A^2\rangle\sim\rho$, with $\rho$ given in eq.~(\ref{rho}).  The
condition $\langle(\partial A)^2\rangle\sim \alpha_s \langle
(A^2)^2\rangle$ translates then into $Q^2\sim \alpha_s \langle
A^2\rangle$, which is eq.~(\ref{Qsaturation}) above. Note that at
saturation, naive perturbation theory breaks down, even though
$\alpha_s(Q_s)$ may be small if $Q_s$ is large: the saturation regime
is a regime of weak coupling, but large density.

Early estimates of $Q_s$ in nucleus-nucleus collisions were given in
\cite{BlaizM1}, and do not differ much from more modern ones
\cite{Muell8}. But, as we have already emphasized, what has changed
dramatically over the last decade is that we do not have only access
to the boundary of the saturation region, we now have a complete
dynamical picture of the saturation region.  Before we turn to a short
introduction to these new developments, let us indicate some expected
characteristics of the saturated regime.

Note first that the saturation momentum increases as the gluon density
increases. This may come from an increase of the gluon structure
function as $x$ decreases. The increase of the density may also come
from the coherent contributions of several nucleons in a nucleus. In
large nuclei, one expects $Q_s^2\propto\alpha_s A^{1/3}$, where $A$ is
the number of nucleons in the nucleus.

There is another feature. Consider the number of partons occupying a
small disk of radius $1/Q_s$ in the transverse plane. This is easily
estimated by combining eqs.~(\ref{rho}) and (\ref{Qsaturation}); one
finds this number to be proportional to $ 1/\alpha_s$.  In such
conditions of large numbers of quanta, classical field approximations
may become relevant to describe the nuclear wave-functions.

\section{Modern formulation:\\ the color glass condensate}

Once one enters the saturated regime the evolution of the parton
distributions can no longer be described by a linear equation such as
the BFKL equation. One of the major breakthrough of the last ten years
is that non linear equations have been obtained which allow us to
follow the evolution of the partonic systems form the dilute regime to
the dense, saturated, regime. These take different, equivalent, forms,
generically referred to as the JIMWLK equation.

The color glass formalism, which provides the most complete physical
picture, relies on the separation of the degrees of freedom in a
high-energy hadron into frozen partons and dynamical fields, as
discussed above. In the original McLerran and Venugopalan model
\cite{McLerV1,McLerV2,McLerV3}, the fast partons are frozen, Lorentz
contracted, color sources flying along the light-cone, and constitute
a density of color charge $\rho(\x_\perp)$. Conversely, the low $x$
partons are described by classical gauge fields $A^\mu(x)$ determined
by solving the Yang-Mills equations with the source given by the
frozen partonic configuration.  An average over all acceptable
configurations must be performed.

The weight of a given configuration is a functional $W_{x_0}[\rho]$ of
the density $\rho$ which depends on the separation scale $x_0$ between
the modes which are described as frozen sources, and the modes which
are described as dynamical fields. As one lowers this separation
scale, more and more modes are included among the frozen sources, and
therefore the functional $W_{x_0}$ evolves with $x_0$ according to a
renormalization group equation
\cite{JalilKLW1,JalilKLW2,Balit1,Kovch1,Kovch3,JalilKMW1,IancuLM1,IancuLM2,Weige1,FerreILM1}.

This evolution equation for   $W_{x_0}[\rho]$ has been
derived in
\cite{JalilKLW1,JalilKLW2,Balit1,%
Kovch1,Kovch3,JalilKMW1,IancuLM1,IancuLM2,Weige1,FerreILM1} and reads
\begin{equation}
\frac{\partial {W_{x_0}[\rho]}}{\partial\ln(1/x_0)}
=\frac{1}{2}\int_{\vec{\x}_\perp,\vec{\y}_\perp}
\frac{\delta}{\delta {\rho_a(\vec{\x}_\perp)}}
{\chi_{ab}(\vec{\x}_\perp,\vec{\y}_\perp)}
\frac{\delta}{\delta\rho_b(\vec{\y}_\perp)}
{W_{x_0}[\rho]}\; ,
\label{eq:JIMWLK}
\end{equation}
where the kernel ${\chi_{ab}(\vec{\x}_\perp,\vec{\y}_\perp)}$ depends
on the color density $\rho$ only via Wilson lines:
\begin{equation}
U(\x_\perp)\equiv {\cal P} \exp\left[ -ig \int_{-\infty}^{+\infty} dz^-
A^+(z^-,\x_\perp)\right]\; .
\end{equation}
Here ${\cal P}$ denotes an ordering along the $x^-$ axis, $A^+$ is
the classical color field of the  hadron moving close to the speed
of light in the $+z$ direction. The field  $A^+$ depends implicitly on the
frozen sources, i.e. on the color charge
density $\rho(\x_\perp)$.

This functional evolution equation can be rewritten as an infinite
hierarchy of equations for correlation functions of the $\rho$'s, or
equivalently of the $U$'s.  For instance, the correlator ${\rm
tr}\big<U^\dagger(\x_\perp)U(\y_\perp)\big>$ of two Wilson lines has
an evolution equation that involves a correlator of four Wilson
lines. If one assumes that this 4-point correlator can be factored
into a product of two 2-point functions, one obtains a closed equation
for the 2-point function, called the Balitsky-Kovchegov
\cite{Balit1,Kovch3} equation:
\begin{eqnarray}
&&\frac{\partial {\rm tr}\big<U^\dagger(\x_\perp)U(\y_\perp)\big>_{x_0}}
{\partial\ln(1/x_0)}
=
-\frac{\alpha_s}{2\pi^2}\int_{\z_\perp}
\frac{(\x_\perp-\y_\perp)^2}{(\x_\perp-\z_\perp)^2(\y_\perp-\z_\perp)^2}
\nonumber\\
&&\times\Big[
N_c {\rm tr}\big<U^\dagger(\x_\perp)U(\y_\perp)\big>_{x_0}
-{\rm tr}\big<U^\dagger(\x_\perp)U(\z_\perp)\big>_{x_0}
{\rm tr}\big<U^\dagger(\z_\perp)U(\y_\perp)\big>_{x_0}
\Big]\; .
\label{eq:BK}
\end{eqnarray}

When the density of color charge $\rho$ is small, one can expand the
Wilson line $U$ in powers of $\rho$.  Eq.~(\ref{eq:BK}) becomes
then\footnote{The same is true of eq.~(\ref{eq:JIMWLK}), because in
this limit the kernel $\chi_{ab}$ becomes quadratic in $\rho$.} a
linear evolution equation for the correlator
$\big<\rho(\x_\perp)\rho(\y_\perp)\big>_{x_0}$, or equivalently for
the unintegrated gluon density, which is nothing but the BFKL
equation.

The opposite limit, at very small $x_0$ (where the color charge
density is large -- far inside the saturation region), also leads to
significant simplifications. Indeed, since the exponent in the Wilson
lines is then a large quantity, one can use a Random Phase
Approximation in which the correlators of Wilson lines are small. The
solution is a Gaussian (albeit non-local in the transverse
coordinates) \cite{IancuIM2}:
\begin{equation}
W_{x_0}[\rho]=\exp\Big[ -\int_{\x_\perp,\y_\perp}
\frac{\rho(\x_\perp)\rho(\y_\perp)}{2\mu^2(\x_\perp-\y_\perp,x_0)}
\Big]\; .
\label{eq:asympt-gaussian}
\end{equation}
This formula is not valid to arbitrarily short distances
$|\x_\perp-\y_\perp|\to 0$, that is the domain of high $Q^2$ physics
controlled by DGLAP evolution.

Like with the BFKL or DGLAP evolution equations, the initial condition
for the evolution is truly a non-perturbative input. One can in
principle try to model it, and then adjust the parameters of the model
to fit experimental data. A simple model is that proposed by McLerran
and Venugopalan, in which the initial $W_{x_0}[\rho]$ is a local
Gaussian:
\begin{equation}
W_{x_0}[\rho]=\exp\Big[ -\int_{\x_\perp}
\frac{\rho(\x_\perp)\rho(\x_\perp)}{2\mu^2}
\Big]\; .
\label{eq:MV-init-cond}
\end{equation}

At this point, one should stress that testing the predictions of the
Color Glass Condensate in principle requires to test both the
properties of the evolution with rapidity and the initial
condition.

It is also important to note that the gaussians of
eqs.~(\ref{eq:asympt-gaussian}) and (\ref{eq:MV-init-cond}) do not
have the same status in the CGC framework: eq.~(\ref{eq:MV-init-cond})
is one particular model for the initial condition at moderately small
$x$, while eq.~(\ref{eq:asympt-gaussian}) is the asymptotic regime
reached at very small $x$ regardless of the initial condition. The
latter is therefore a property of the small $x$ evolution itself. The
MV model requires an infrared cutoff at the scale $\Lambda_{_{QCD}}$.
This is because assuming a truly local Gaussian distribution ignores
the fact that color neutralization occurs on distance scales smaller
than the nucleon size ($\sim \Lambda_{_{QCD}}^{-1}$): two $\rho$'s can
only be uncorrelated if they are at transverse coordinates separated
by at least the distance scale of color neutralization.  In the
asymptotic solution of eq.~(\ref{eq:asympt-gaussian}), there is no
infrared problem because the color neutralization is built in the
dependence of $\mu^2(\x_\perp-\y_\perp,x_0)$ on the transverse
coordinates. Color neutralization in fact occurs on distance scales of
the order of $Q_s^{-1}(x_0)$ \cite{IancuIM2}.  This is the physical
origin of the universality of the saturated regime.

\section{Phenomenological predictions}
\subsection{Geometrical scaling}
\begin{figure}[htbp]
\begin{center}
\resizebox*{!}{4cm}{\includegraphics{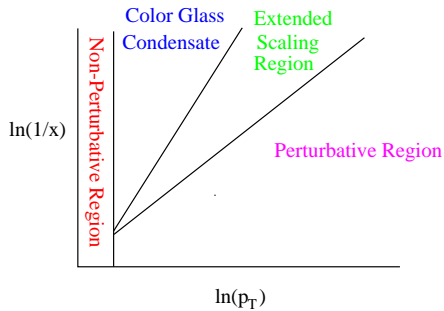}}
\end{center}
\caption{\label{fig:scaling-domain} The scaling region in the
$(\ln(1/x),\ln(p_\perp))$ plane.}
\end{figure}

In the $(\ln(1/x),\ln(p_\perp))$ plane, the saturation domain is
defined by the condition $\Lambda_{_{QCD}}\ll p_\perp \lesssim
Q_s(x)$. In this region, one expects that observables (made
dimensionless by scaling out the appropriate power of $p_\perp$)
depending a priori on both $x$ and $p_\perp$ scale like functions of
the unique variable $p_\perp/Q_s(x)$. This property is called
``geometrical scaling'' in the literature. It has been shown
by Iancu, Itakura and McLerran in \cite{IancuIM1} that this scaling
property holds in a much larger domain, whose upper bound is
$p_\perp \lesssim Q_s(x)^2/\Lambda_{_{QCD}}$. This region has been
labeled ``extended scaling region'' in figure
\ref{fig:scaling-domain}.
Technically, the reason why this scaling survives outside of the
saturation domain is that the (linear) BFKL equation that controls the
evolution in $x$ when $p_\perp \gtrsim Q_s(x)$ preserves for a while
the scaling properties inherited from the physics of the saturation
region (which affects the BFKL evolution via a boundary condition at
$p_\perp=Q_s(x)$).

\begin{figure}[htbp]
\begin{center}
\resizebox*{!}{6cm}{\includegraphics{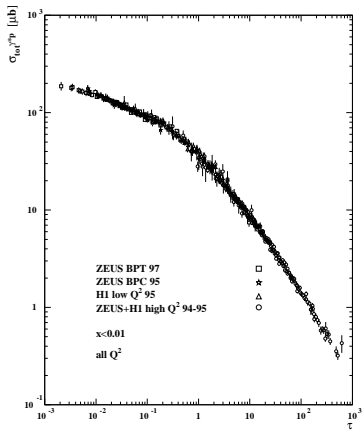}}
\end{center}
\caption{\label{fig:scaling-F2} The $\gamma^*p$ cross-section at HERA,
plotted against $\tau\equiv Q^2/Q_s^2(x)$.}
\end{figure}

Such a scaling has been searched for in the data of the DIS experiments at
HERA, and it turns out that one can indeed represent all the small $x$
($x<10^{-2}$) data points for the structure function $F_2(x,Q^2)$ as a
single function of the variable $\tau\equiv Q^2/Q_s^2(x)$ with
\begin{equation}
   Q_s^2(x)\equiv Q_0^2 \left(\frac{x_0}{x}\right)^\lambda\; .
\label{eq:Qs-fit}
\end{equation}
This  is illustrated by the plot of figure
\ref{fig:scaling-F2}.
The parameter $Q_0$ is set to $1$~GeV, while $x_0\approx 3.10^{-4}$ and
$\lambda\approx 0.29$ are determined through a fit  to the data
\cite{StastGK1}. By
studying BFKL evolution to next-to-leading order,
Triantafyllopoulos has obtained a value of
$\lambda$ in good agreement with that observed \cite{Trian1}.

\subsection{Structure functions and diffraction in DIS}

In an
appropriate frame, and at leading order in $\alpha_s$, the cross
section for the interaction of a virtual photon with a proton
takes the following factorized form:
\begin{equation}
\sigma_{\gamma^* p}(x,Q^2)=\int\limits_0^1 dz\int d^2\r_\perp
\left|\psi(Q^2,z,\r_\perp)\right|^2
\sigma_{\rm dipole}(x,\r_\perp)\; .
\label{eq:fact-F2}
\end{equation}
In this formula, $\psi(Q^2,z,\r_\perp)$ is the Fock component of the
virtual photon light-cone wave function that corresponds to a
$q\bar{q}$ dipole of size $\r_\perp$; it depends on the invariant mass
$Q^2$ of the photon, on the transverse size $\r_\perp$ of the
$q\bar{q}$ dipole, and on the fraction $z$ of the photon longitudinal
momentum taken by the quark.  The other factor in this formula,
$\sigma_{\rm dipole}(x,\r_\perp)$, is the total dipole-proton
cross-section.  It can be expressed in terms of a correlator of Wilson
lines:
\begin{equation}
\sigma_{\rm dipole}(x,\r_\perp)=\frac{2}{N_c}
\int d^2\X_\perp
{\rm tr} \,\left[1-\left<U(\X_\perp+\frac{\r_\perp}{2})
U^\dagger(\X_\perp-\frac{\r_\perp}{2})\right>\right]\; ,
\label{eq:dipole-cs}
\end{equation}
where the average is taken over the color field of the proton.

Several models for this dipole cross-section have been used in the
literature in order to fit HERA data. Golec-Biernat and W\"usthoff
have used a very simple parameterization \cite{GolecW1,GolecW2}
\begin{equation}
\sigma_{\rm dipole}(x,\r_\perp)=\sigma_0 \left[
1-e^{-\frac{1}{4}Q_s^2(x)r_\perp^2}
\right],
\end{equation}
which has led to good results for describing the data at $x<10^{-2}$
and moderate $Q^2$. In this formula, the scale $Q_s(x)$ was taken to
be of the form given in eq.~(\ref{eq:Qs-fit}). This model fails
however at large $Q^2$. This aspect was improved in \cite{BarteGK1},
where the dipole cross-section is parameterized in a way that
reproduces pQCD for small dipoles. Note that these approaches, even if
they are inspired by saturation physics, do not derive the dipole
cross-section from first principles. Recently, Iancu, Itakura and
Munier \cite{IancuIM3} derived an expression of the dipole
cross-section from the color glass condensate framework, and obtained
a good fit of HERA data with a small number of free parameters. An
equally good fit has been obtained by Gotsman, Levin, Lublinsky and
Maor who derived the $x$ dependence of the dipole cross-section by
solving numerically the BK equation and including DGLAP corrections 
\cite{GotsmLLM1}.

One can express in a similarly factorized form the diffractive
$\gamma^*p$ cross-section, where the exchange between the dipole and
the target is color singlet. In fact, since the dipole cross-section
involved in eqs.~(\ref{eq:fact-F2}) and (\ref{eq:dipole-cs}) is the
total cross-section, one can use the  optical theorem to deduce from
it the forward elastic scattering amplitude:
\begin{equation}
{\cal M}_{\rm elastic}(x,\r_\perp,t=0)=\frac{i}{N_c}
\int d^2\X_\perp
{\rm tr} \,\left[1-\left<U(\X_\perp+\frac{\r_\perp}{2})
U^\dagger(\X_\perp-\frac{\r_\perp}{2})\right>\right]\; ,
\end{equation}
assuming that this amplitude is predominantly imaginary.
This can then be  used to  obtain for the diffractive DIS
cross-section at $t=0$ an
expression involving  the {\it square} of the total dipole-proton
cross-section\footnote{Another way in which this relation is stated in
    the literature is that in order to obtain the diffractive part of a
    cross-section, one should average the scattering amplitude over the
    configurations of the frozen sources {\it before} squaring the
    amplitude.}:
\begin{equation}
\left.\frac{d\sigma^{^D}_{\gamma^* p}(x,Q^2)}{dt}\right|_{t=0}
=\frac{1}{16\pi}\int\limits_0^1 dz\int d^2\r_\perp
\left|\psi(Q^2,z,\r_\perp)\right|^2
\sigma_{\rm dipole}^2(x,Q^2)\; .
\end{equation}
By using in this formula the total dipole cross-section fitted to the
inclusive $F_2$, one obtains a good description of the measured
diffractive DIS cross-section \cite{GolecW1,GolecW2}. Note however that
keeping only the $q\bar{q}$ component of the photon wave
function is a good approximation only when the invariant mass of the
diffractively produced object is not too large. For large masses, one
needs to consider also the $q\bar{q}g$ component.

\subsection{Particle multiplicities at RHIC}

In the framework of saturation models, predictions have been made for
global observables like the total number of produced particles per
unit of rapidity $dN/d\eta$. This can be studied as a function of the
collision energy, the centrality of the collision, and the rapidity.

\begin{figure}[htbp]
\begin{center}
\resizebox*{!}{6cm}{\includegraphics{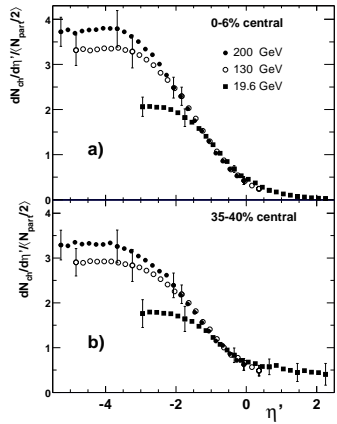}}
\end{center}
\caption{\label{fig:limiting-frag} Particle multiplicity at RHIC for
    different collision energies, plotted against the rapidity of the
    produced particle relative to the rapidity of the beam
    $\eta^\prime\equiv\eta-\eta_{\rm beam}$ \cite{Backa1}.}
\end{figure}
The Color Glass Condensate may provide a dynamical justification
rooted in QCD for many general features of particles production in
hadronic interactions. This may be the case in particular for the
phenomenon of ``limiting fragmentation", i.e., the expectation that
the rapidity distribution in the fragmentation region becomes
independent of the collision energy at high energy. Evidence for such
a behavior has indeed been observed by the PHOBOS collaboration, as
is illustrated in figure \ref{fig:limiting-frag}. Here, one has
plotted the rapidity distribution $dN/d\eta^\prime$ of produced
particles in Au-Au collisions as a function of the rapidity relative
to the incoming beam rapidity $\eta^\prime\equiv \eta-\eta_{\rm
   beam}$.  When one increases the beam energy, one sees that
$dN/d\eta^\prime$ is the same at large $\eta^\prime$ for all energies.
The particles produced at some given (large) energy are those that
would have been produced at lower energies plus new particles produced
at a lower rapidity. An early interpretation of such a phenomenon in
the language of the CGC has been proposed by Jalilian-Marian
\cite{Jalil3}. The interpretation suggested there is that in the
fragmentation region of one of the nuclei ($\eta^\prime$ around zero),
this nucleus is a dilute partonic system while the other nucleus is in
a saturated state. The rise in multiplicity when one decreases
$\eta^\prime$ is then attributed to the growth of the parton
distribution in the dilute nucleus.  One should emphasize however that
no detailed quantitative analysis of the data presented in figure
\ref{fig:limiting-frag} has yet been done in this framework.

In comparing more quantitative predictions of the color glass
formalism for $dN/d\eta$ to experimental data, it is important to keep
the following points in mind. First, the color glass condensate can
only predict the distribution of initial gluons, set free typically at
a proper time $\tau\sim Q_s^{-1}$. Between this early stage and the
final chemical freeze-out, the system undergoes several non-trivial
steps: kinetic and chemical equilibration (possibly with additional
parton production), hadronization, etc., which are most often ignored
in calculations based on the color glass condensate. Secondly, even
the calculation of the initial gluon production is a highly
non-trivial task. It involves, in principle, solving the classical
Yang-Mills equations in the presence of the color densities describing
the distribution of frozen sources in both projectiles. This has been
achieved analytically only for proton-nucleus collisions, when the
color charge density inside the proton is assumed to be weak
\cite{KovchM3,DumitM1,BlaizGV1}. In the case of nucleus-nucleus
collisions, two kinds of calculations have been performed:

(i) {\it Ab initio} numerical calculations
\cite{KrasnV1,KrasnV2,KrasnNV2,Lappi1} that solve exactly the
Yang-Mills equations for $AA$ collisions. The average over the color
charge densities in the projectiles is performed by assuming that the
distributions $W_{x_0}[\rho]$ are given by the MV model, i.e. by the
gaussian of eq.~(\ref{eq:MV-init-cond}). Quantum evolution effects are
therefore not included. Note that when one calculates $dN/d\eta$ at
$\eta=0$, the typical $x$ probed in the wave function of the nuclei is
about $x\sim 10^{-2}$. Such calculations can therefore be used to test
the validity of the gaussian model\footnote{Since the gluon
    multiplicity depends only on the correlator
    $\big<A^\mu(x)A^\nu(y)\big>$, only one moment of the distribution
    $W_{x_0}[\rho]$ is tested by this observable. One could certainly
    imagine different models for this distribution that give the same
    value for this particular correlator.} for $W_{x_0}[\rho]$ at
$x_0\sim 10^{-2}$ as well as the non-linear dynamics of the classical
field.

(ii) Approximate analytical calculations of the initial gluon spectrum
\cite{KharzL1}. These calculations assume $k_\perp$-factorization,
although such a property is so far unproven for the collision of two
dense projectiles.
\begin{figure}[htbp]
\centerline{\hfil
\resizebox*{!}{4.5cm}{\includegraphics{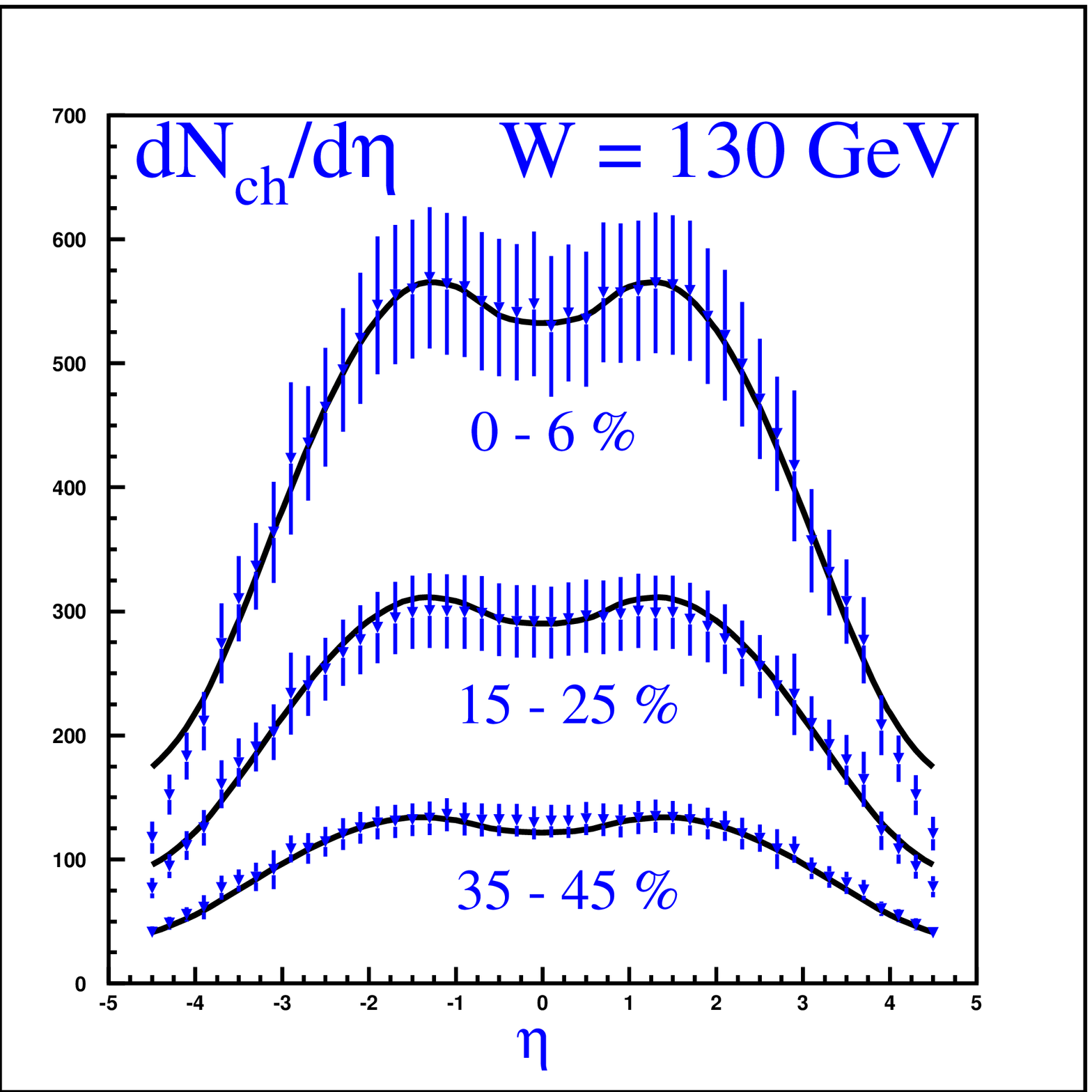}}
\hfil\hglue 5mm\hfil
\resizebox*{!}{4.5cm}{\includegraphics{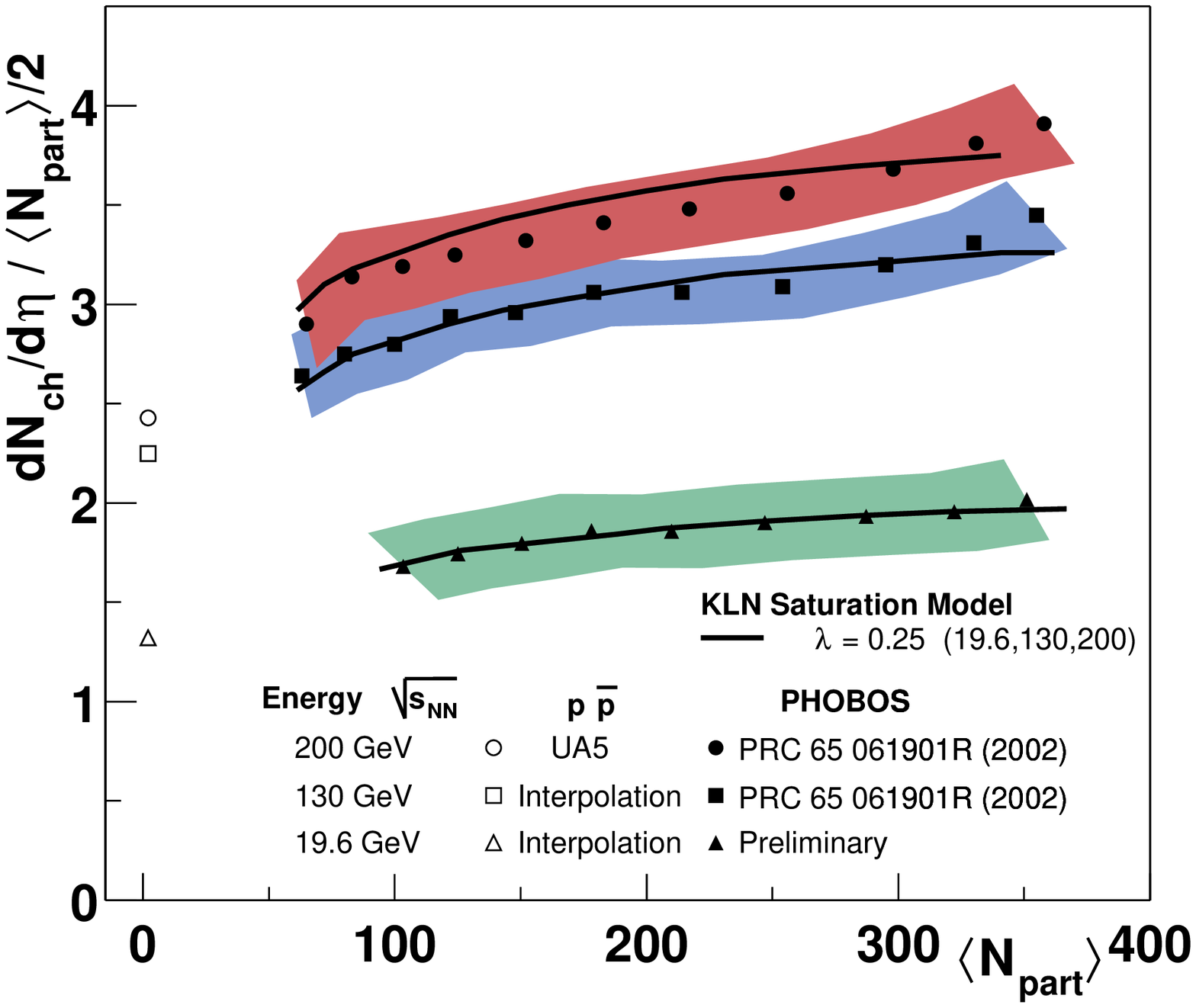}}
\hfil}
\caption{\label{fig:dndy-AA} The number of particles per unit of
rapidity in $AA$ collisions, as a function of rapidity for different
centralities (left plot) and as a function of centrality at different
energies (right plot). From \cite{KharzL1} .}
\end{figure}
The non-integrated gluon distribution in the nuclei is taken to be of
the form $\varphi(k_\perp)\sim 1/\alpha_s$ for $k_\perp\ll Q_s(x)$ and
$\varphi(k_\perp)\sim Q_s^2(x)/k_\perp^2$ for $k_\perp\gg Q_s(x)$
(saturation appears through the fact that the gluon distribution does
not diverge like $k_\perp^{-2}$ at small $k_\perp$).  The rapidity
dependence in this model is governed by that of $Q_s(x)$,
eq.~(\ref{eq:Qs-fit}), where the exponent $\lambda$ can be taken from
the study of the scaling properties in the HERA data. The overall
normalization factor, as well as the value of $Q_s$ at a certain fixed
energy, were fitted on RHIC data at $\sqrt{s}=130$~GeV. A prediction
was then made in \cite{KharzL1} for the value of $dN/d\eta$ at
$\sqrt{s}=200$~GeV, which is in good agreement with experimental
results, as illustrated in figure \ref{fig:dndy-AA}. One should also
mention the fact that the value of $Q_s^2(x)$ one needs in this
approach in order to reproduce the measured $dN/d\eta$, of the order
of $2$~GeV${}^2$, is relatively large compared to $\Lambda_{_{QCD}}$;
this may perhaps be taken as an encouraging indication for the
validity of the overall picture.

Note that the residual dependence on the number of participants in the
quantity $N_{\rm part}^{-1}dN/d\eta$ calculated by Kharzeev, Levin and
Nardi, comes entirely from the scale dependence of the strong coupling
constant, as follows:
\begin{equation}
\frac{1}{N_{\rm part}}\frac{dN}{d\eta}\sim
\frac{1}{\alpha_s(Q_s^2)}\sim \ln(Q_s^2/\Lambda_{_{QCD}}^2)\; .
\end{equation}
However, strictly speaking, since the gluon multiplicity is obtained
by solving the {\it classical} Yang-Mills equations, there is no
running of $\alpha_s$ at this level of approximation. Certainly the
running of $\alpha_s$ will come together with next-to-leading-order
corrections, but it is at this point an ad hoc prescription. Since
other interpretations of these same data, based on soft physics, are
possible \cite{CapelS1}, it is unclear whether what we are ``seeing''
in the right hand panel of figure \ref{fig:dndy-AA} is indeed the
running of $\alpha_s(Q_s^2)$ induced by the variation of the
saturation scale with centrality.

\begin{figure}[htbp]
\centerline{\hfil
\resizebox*{!}{4.5cm}{\includegraphics{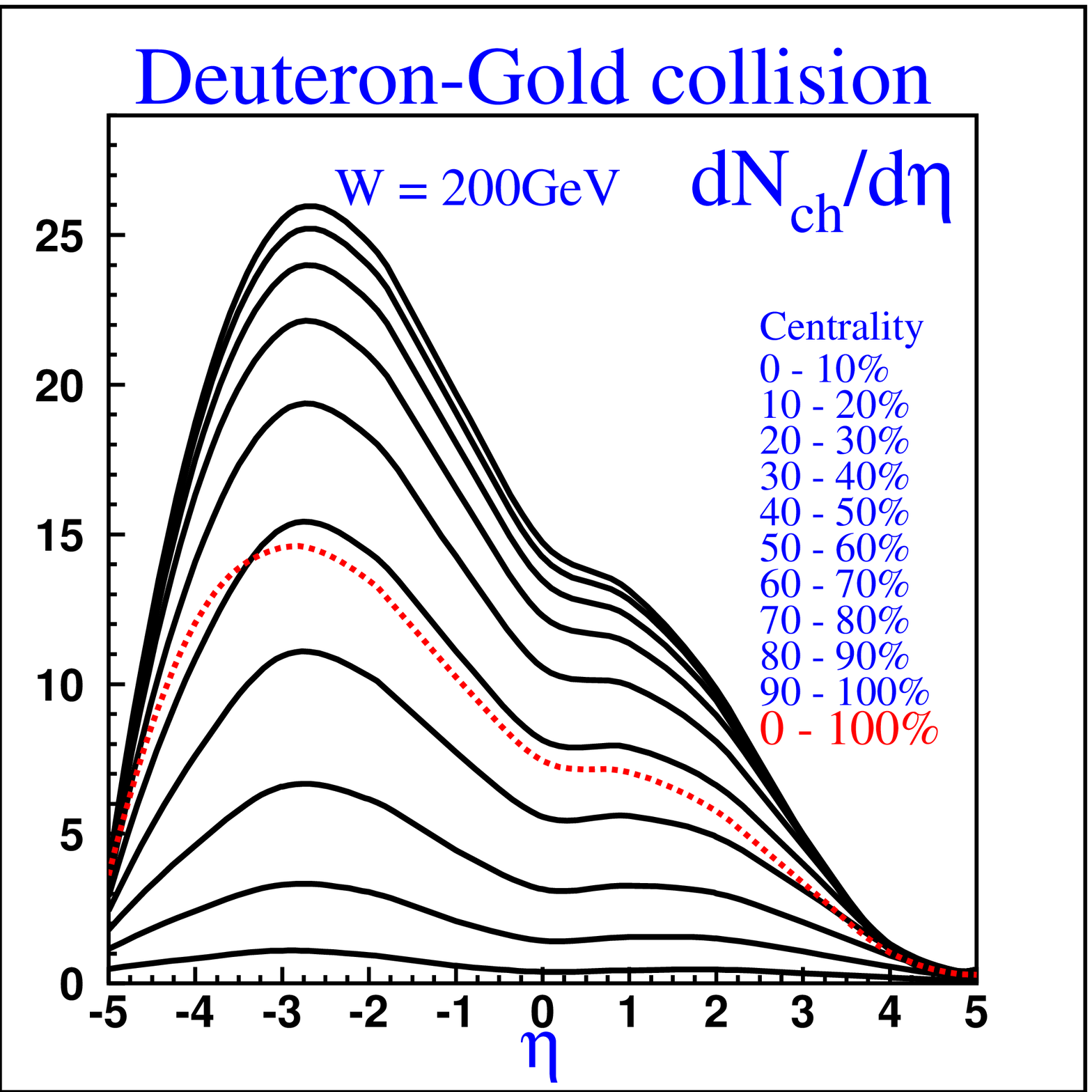}}
\hfil\hglue 5mm\hfil
\resizebox*{!}{4.5cm}{\includegraphics{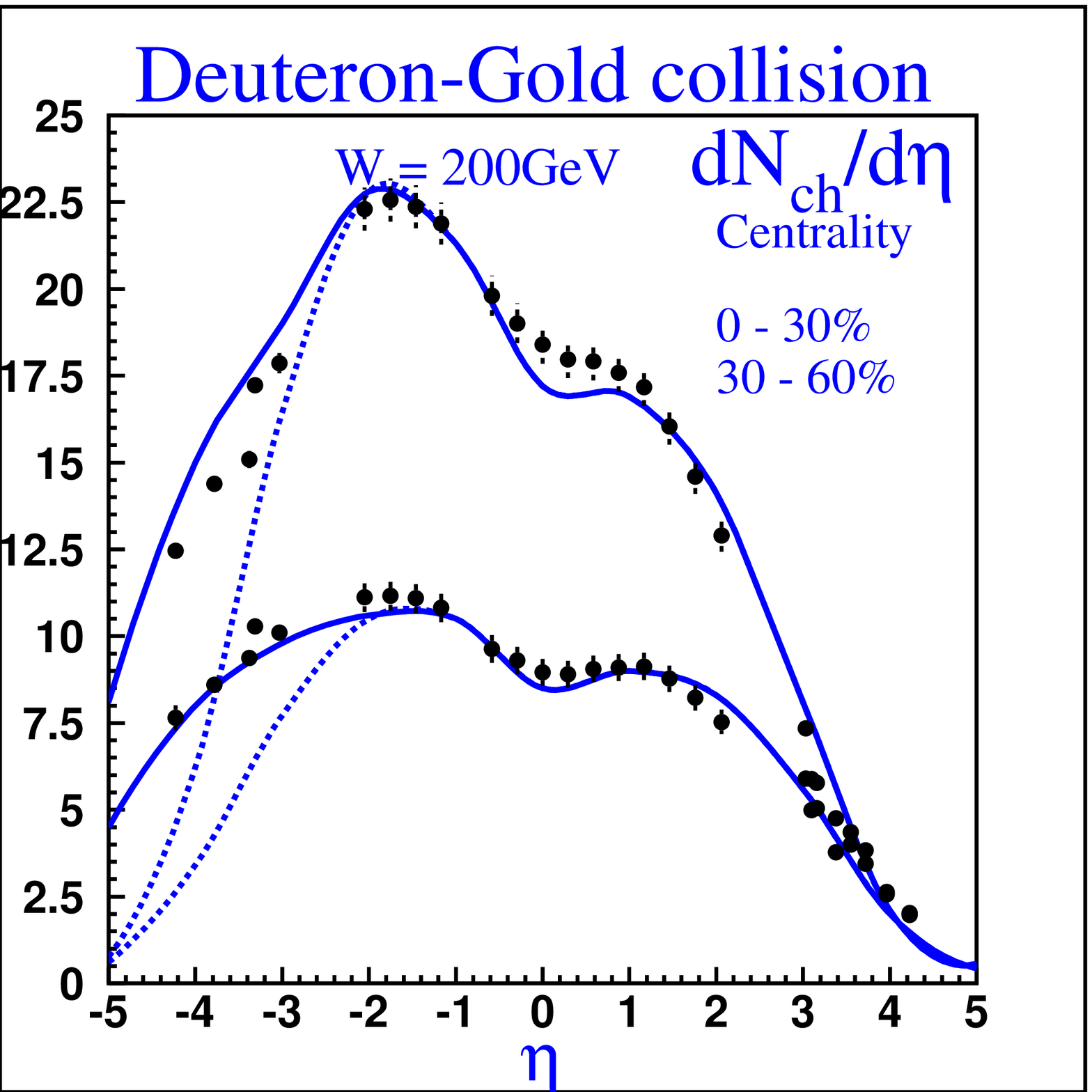}}
\hfil}
\caption{\label{fig:dndy-dA} The number of particles per unit of
rapidity in $dA$ collisions, as predicted in \cite{KharzLN2}.}
\end{figure}
Such analysis of multiplicity distributions were extended   to
the case of deuteron-nucleus collisions \cite{KharzLN2}, with results in fair
agreement with RHIC data\footnote{In the first version of this study,
there was a discrepancy with the measured multiplicities, which seems
to have been pinned down to a problem with the Glauber determination
of the number of participants.} as illustrated in figure
\ref{fig:dndy-dA}.

\subsection{Spectra of produced particles}
\subsubsection{$p_\perp$ dependence}

The saturation regime is characterized by a single scale, the
saturation momentum $Q_s$, and as was the case for DIS, one may look
for scaling laws in various observables.  An attempt has been made to
identify such a scaling in the transverse mass distributions of
produced particles \cite{SchafKMV1}: it has been found that the
spectra for different species of particles are well represented by a
single function of $m_\perp/Q_s$. The centrality dependence of the
scaling parameter $Q_s^2$ was found in rough agreement with $Q_s^2\sim
N_{\rm part}^{1/3}\sim A^{1/3}$. It should be emphasized however,
that the hadrons whose spectra are measured have certainly undergone
many reinteractions, so that their spectra are unlikely to reflect
directly the initial momentum distribution of the color glass. Thus
the interpretation of the $m_\perp$ scaling in the saturation picture is
unconvincing.

In the case of deuteron-gold collisions, one does not expect final
state interactions to play a dominant role. All the medium effects
responsible for the difference between $pA$ and $pp$ collisions may in
fact be taken into account in the color glass condensate. The produced
partons hadronize then by the same mechanisms as in the vacuum, and
their spectrum can be calculated by convoluting the partonic
cross-section with the usual fragmentation functions. Moreover, since
the calculation of the gluon spectrum only involves the correlator
$\big<U(0)U^\dagger(\x_\perp)\big>$
\cite{DumitM1,DumitJ1,DumitJ2,BlaizGV1}, one can in principle
determine it from the dipole cross-section used in DIS \cite{GelisJ3}.
A numerical calculation of the spectrum of hadrons produced in dA
collisions has been performed in this approach in \cite{Jalil1}, with
results in fairly good agreement with the spectrum measured at RHIC in
$dA$ collisions.

The situation is far more complicated in the case of nucleus-nucleus
collisions. Indeed, the gluons which emerge from the color glass
shortly after the initial impact (typically at a proper time of the
order of $Q_s^{-1}\sim 0.2$~fm) will interact strongly and may form a
hot and dense medium (a quark-gluon plasma?).  These interactions will
presumably modify the gluon spectrum: the momentum distribution will
become more isotropic and the hard tail of the spectrum will be
suppressed by parton energy loss.  These additional effects are not
taken into account in the color glass, which merely provides the
initial condition for the subsequent evolution of the system.

By assuming that local thermalization is achieved quickly, one can use
hydrodynamics to describe this evolution. One may adjust the initial
conditions so that averaged quantities like the local energy density
match those predicted in saturation models. Such a strategy has been
used by Hirano and Nara \cite{HiranN1} as well as Eskola, Niemi,
Ruuskanen and Rasanen \cite{EskolNRR2,EskolNRR1}, who predicted hadron
spectra in good agreement with the observed ones for $p_\perp \lesssim
2$~GeV.  One should however keep in mind that this kind of test is by
construction only sensitive to integrated quantities predicted by the
CGC rather than to the detailed shape of the spectrum: indeed, the
very assumption that thermalization occurs means that the system has
lost memory of the initial momentum distribution except for the local
energy density.

This remark implies that $dA$ collisions are certainly much better
suited than $AA$ collisions in order to test the predictions of the
color glass condensate, because the details of the system at early
times are carried out to the final state almost unaltered. One should
therefore be able to perform more direct tests of the CGC ideas
in the context of $dA$ collisions.

\subsubsection{Anisotropy}
Another interesting quantity to look at is the so-called ``elliptic
flow'', which signals the existence of a significant pressure in the
transverse direction \cite{Ollit1}, that converts the original spatial
anisotropy into an anisotropy in momentum space.

\begin{figure}[htbp]
\centerline{\hfil\
\resizebox*{!}{5cm}{\includegraphics{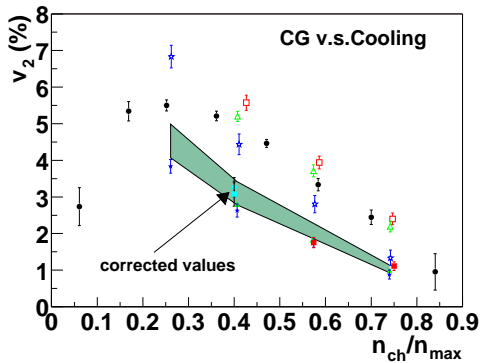}}
\hfil}
\caption{\label{fig:v2} The centrality dependence of $v_2$ from the
classical gluon production (green band). The black filled dots are
STAR data.}
\end{figure}

This has been estimated by Krasnitz, Nara and Venugopalan
\cite{KrasnNV3} from the numerical solution of the classical
Yang-Mills equations. The growth of the anisotropy with time is
similar to what happens in hydrodynamical models, \cite{KrasnNV3} and
the full $v_2$ is attained after a time of the order of the size $R$
of the system.  The $v_2$ obtained at late times in the color glass
condensate framework is plotted in figure
\ref{fig:v2}.
One can see that  although it is not in
   quantitative agreement with the data, it has the correct order of
magnitude and reproduces qualitatively the dependence on the
multiplicity (i.e. on the centrality). However, the dependence of
$v_2$ on the transverse momentum is in clear disagreement with the
RHIC data: the color glass condensate predicts that $v_2$ has a
maximum at small momentum and then falls to zero at large momentum,
while the measured $v_2$ increases monotonously up $p_\perp\sim 2$~GeV
and remains almost constant at higher momenta. Note   that  the 
description of the system in terms of classical
    fields breaks down for times that are large compared to 
$Q_s^{-1}$, that is for times needed to establish the $v_2$. It is 
nevertheless instructive to see that this model, which does not
    rely on the assumption of thermalization, produces a $v_2$ of
    comparable magnitude to the $v_2$ obtained in hydrodynamical
    expansion over similar time-scales.

\subsection{Ratios of spectra: $R_{AA}$, $R_{dA}$}
The comparison of the $p_\perp$ spectra in $AA$ or $dA$ to the
corresponding ones in $pp$ collisions is usually done through the
following ratio:
\begin{eqnarray}
R_{AA}\equiv
\frac{\left.\frac{dN}{dyd^2\p_\perp}\right|_{AA}}
{N_{\rm coll}\left.\frac{dN}{dyd^2\p_\perp}\right|_{pp}}
\; ,\quad
R_{dA}\equiv
\frac{\left.\frac{dN}{dyd^2\p_\perp}\right|_{dA}}
{N_{\rm coll}\left.\frac{dN}{dyd^2\p_\perp}\right|_{pp}}\; ,
\end{eqnarray}
where $N_{\rm coll}$ is the number of binary collisions. For the
production of high $p_\perp$ particles,  expected to scale
like the number of binary collisions, these ratios should be unity.
These ratios have been measured at RHIC both for $AA$ and $dA$
collisions.

\begin{figure}[htbp]
\centerline{\hfil\
\resizebox*{!}{3.3cm}{\includegraphics{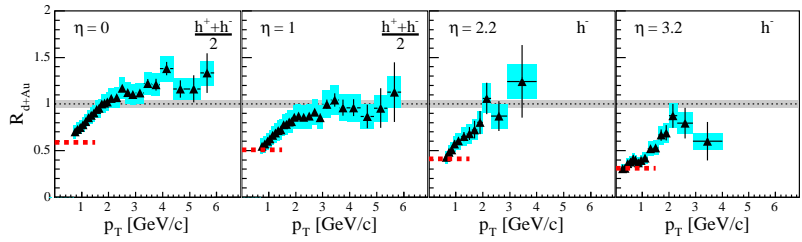}}
\hfil}
\caption{\label{fig:brahms} The ratio $R_{dA}$ measured at RHIC by
    the BRAHMS collaboration at several rapidities (positive rapidities
    correspond to the fragmentation region of the deuteron)
    \cite{Arsena1}.}
\end{figure}

In $AA$ collisions at central rapidity ($\eta=0$), the measured
$R_{AA}$ is much smaller than one (around $0.2$) in central
collisions, even at $p_\perp$'s as high as $10$~GeV, and approaches
one from below as one goes to more and more peripheral collisions.
However, at the values of $x$ probed by gluon production at
mid-rapidity ($x\sim 10^{-2}$ for $p_\perp\sim 1$~GeV), where one
expects the distribution $W_{x_0}[\rho]$ to be given by the
McLerran-Venugopalan model (eq.~(\ref{eq:MV-init-cond})), the $R_{AA}$
predicted in the color glass condensate at a time of the order of
$Q_s^{-1}$ has a maximum larger than $1$ at small $p_\perp$ and then
goes to one from above as $p_\perp$ increases \cite{JalilNV1}.  This
maximum of $R_{AA}$ in the MV model is interpreted as a manifestation
of the multiple (elastic) rescatterings of the produced gluon, which
tend to redistribute the $p_\perp$ spectrum by depleting the small
$p_\perp$'s and enhancing the high $p_\perp$'s (Cronin effect). The
observed suppression of $R_{AA}$ therefore requires final state
effects, and is naturally interpreted as energy loss
(\cite{Bjork3,WangG1,GyulaW1,BaierDPS1,BaierDMPS2}, see also
\cite{GyulaVWZ1} for a recent review) of the high $p_\perp$ partons as
they go through the dense medium formed in $AA$ collisions.

\begin{figure}[htbp]
    \centerline{\hfil\
      \resizebox*{!}{7cm}{\includegraphics{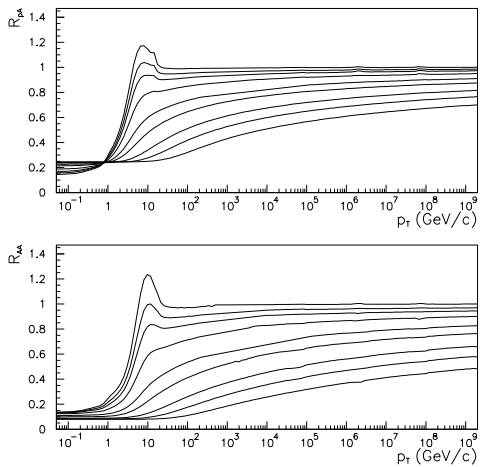}} \hfil}
\caption{\label{fig:armesto} Evolution of the ratios $R_{AA}$ and
    $R_{dA}$ with rapidity in the color glass condensate framework.  The
    initial condition is the McLerran-Venugopalan model. Note that the
    calculation of $R_{AA}$ assumed $k_\perp$-factorization.}
\end{figure}

At much smaller values of $x$, which can be probed by looking at
particle production at large rapidity, one expects the distribution
$W_{x_0}[\rho]$ to reach the asymptotic form of
eq.~(\ref{eq:asympt-gaussian}). It has been found that with such a
distribution of color sources, the color glass condensate leads to a
suppression of the ratios $R_{AA}$ or $R_{dA}$ \cite{KharzLM1} (see
the plots of figure \ref{fig:armesto}, taken from \cite{AlbacAKSW1}).
However, in order to disentangle this suppression which is an initial
state effect (because it comes from the ``wave function'' of the
projectiles) from the energy loss, one needs an experiment in which
one has no significant final state effects, like $dA$ collisions. The
ratio $R_{dA}$ measured by the BRAHMS experiment is shown in figure
\ref{fig:brahms} \cite{Arsena1}. One can see at $\eta=0$ a very
different behavior than for $AA$ collisions: the ratio $R_{dA}$ has a
maximum above one and seems to go to one from above at high $p_\perp$.
As one increases the rapidity of the observed particles, one can see
the ratio $R_{dA}$ drop very fast and eventually become consistently
smaller than one.  Such a behavior with rapidity was predicted in
\cite{AlbacAKSW1} (see the figure \ref{fig:armesto}), where the
evolution of the ratio $R_{dA}$ was evaluated by evolving the
unintegrated gluon distribution in a nucleus using the
Balitsky-Kovchegov equation, starting from the McLerran-Venugopalan
model at large $x$.
Similar qualitative results were also obtained in \cite{BaierSW1}, in
\cite{KharzKT1} within a toy model for the gluon distribution inspired
from the color glass condensate, and in \cite{BlaizGV1} where the two
distributions of frozen sources given in
eqs.~(\ref{eq:asympt-gaussian}) and (\ref{eq:MV-init-cond}) were used
in order to compute the ratio $R_{dA}$.  The onset of the suppression
of the ratio $R_{dA}$ as one increases the rapidity was studied
analytically in \cite{IancuIT2}, where it was shown that the rapid
suppression seems to be due to the fact that the gluon distribution
evolves faster in a proton than it does in a nucleus, because the
nucleus reaches the saturation regime earlier. Recently, a more
quantitative study of this effect has been performed in
\cite{KharzKT2}. So far, the color glass condensate is the only
framework in which one reproduces, at least qualitatively, both the
the Cronin effect at central rapidities and its suppression at forward
rapidities.

\section{Discussion and outlook}

Considerable theoretical progress has been made over the last ten
years in understanding the physics that govern the wave function of a
hadron at high energy. Not only has one acquired a fairly intuitive
picture that is universally applicable to all hadrons at high energy,
but an operational framework has been developed in which many
phenomena can be described quantitatively. One is now in principle
able to study and make predictions in the entire $x,k_\perp$ plane
(with the exception of the truly non-perturbative region
$k_\perp\lesssim \Lambda_{_{QCD}}$), with non-perturbative physics
entering only as an initial condition for otherwise known evolution
equations.

With the realization that the characteristic scale that governs high
energy hadronic interactions, the saturation momentum $Q_s$, is
enhanced by the size of the projectiles in collisions involving
nuclei, RHIC can be viewed as a place of choice in order to test these
ideas and confront them to data. And indeed the phenomenology based on
saturation physics has been quite successful in describing what is
observed at RHIC. Whenever relevant comparisons with data can be made
(global observables in $AA$ collisions, spectra in $dA$ collisions),
the tests are successful. In cases of marked disagreement, like with
the elliptic flow, one understands that the discrepancy comes from
using the color glass picture beyond its domain of applicability.

One may ask whether the predictions which have been tested so far are
distinctive features of the color glass condensate that cannot be
reasonably explained by other models. Naturally, the more precise the
question, the more discriminatory the answer. Because in $AA$
collisions the final state interactions are important, only global
observables are preserved from early times to the final state.
Therefore $AA$ collisions are in general not ideal for direct tests of
the color glass condensate (many global features of $AA$ collisions
are present in any reasonable model of hadronic interactions, and thus
not characteristic of the color glass condensate).  Collisions
involving a small projectile, like $dA$ collisions, where effects of
final state interactions can be minimized, at least in some
kinematical domains, are much better suited.  And indeed the
suppression of the ratio $R_{dA}$ at forward rapidities could turn out
to be such a distinctive measurement, since no other model has been
able so far to provide a natural explanation of what is observed. This
is also one instance where non trivial quantum evolution could be
playing a significant role.  What is missing for this to become a real
test is a detailed calculation of this effect that would go beyond the
many qualitative approaches performed so far. Clearly, doing such
calculations, as well as getting more precise data, is of utmost
importance.

  From comparisons with data, one has also learned that final state
interactions are generally very important in $AA$ collisions, and many
observables do not simply reflect the properties of the partonic
system produced in the early stages. The color glass condensate only
provides the initial condition for a subsequent evolution of the
system leading possibly to the formation of a quark-gluon
plasma. Understanding whether and how thermalization happens presents
interesting theoretical challenges (see
\cite{BaierMSS2,RomatS1,MrowcRS1,ArnolLM1,BergeBW1} for some recent
works on the subject) and is of utmost importance for giving a solid
theoretical basis to present descriptions of $AA$ collisions.

As a final remark, oriented towards the future, one may note that many
of the phenomena uncovered at RHIC, should become more clearly visible
at the LHC. There, with center of mass energies of $5.5~$TeV for $AA$
collisions, the typical value of $x$ at mid-rapidity will be about
$5.10^{-4}$ and values as small as $10^{-5}$ could be reached at
forward rapidities. This corresponds roughly to values of $Q_s^2$
ranging from $5~$GeV${}^2$ to $14$~GeV${}^2$. Such large values of the
saturation momentum make the coupling constant $\alpha_s$ smaller than
at RHIC, giving firmer grounds to the weak coupling expansion used in
the color glass condensate framework.  A corollary of this is that the
gluon occupation number in the saturation region will be larger,
making the classical description also more justified.

\section*{Acknowledgements}
We would like to thank E. Iancu and A.H. Mueller for their careful
reading of this manuscript.


\end{document}